\documentclass{article}
\newfont{\eaddfntsmall}{phvr8t at 10pt}

\usepackage{amsmath,amsthm,amsfonts,amssymb,graphicx}
\usepackage[utf8x]{inputenc}
\usepackage[T1]{fontenc}

\usepackage{url,xspace}
\usepackage{algorithm,%
algpseudocode%
}
\usepackage{float}

\usepackage{rotating}
\usepackage{multicol}

\usepackage{stmaryrd}
\def\oc{\left\llbracket}
\def\fc{\right\rrbracket}

\usepackage{mathtools}
\def\transpo#1{\prescript{\top\mkern -2mu}{}{#1}}
\DeclareMathOperator{\Zb}{\mathbf{Z}}
\DeclareMathOperator{\Xb}{\mathbf{X}}
\DeclareMathOperator{\Yb}{\mathbf{Y}}
\DeclareMathOperator{\xb}{\mathbf{x}}
\DeclareMathOperator{\yb}{\mathbf{y}}

\newcommand{\coo}{\texttt{COO}\xspace}
\newcommand{\coos}{\texttt{COO\_S}\xspace}
\newcommand{\csr}{\texttt{CSR}\xspace}
\newcommand{\dia}{\texttt{DIA}\xspace}
\renewcommand{\ell}{\texttt{ELL}\xspace}
\newcommand{\ellr}{\texttt{ELL\_R}\xspace}
\newcommand{\nbnz}{\texttt{nbnz}\xspace}
\newcommand{\data}{\texttt{data}\xspace}
\newcommand{\rowid}{\texttt{rowid}\xspace}
\newcommand{\colid}{\texttt{colid}\xspace}
\newcommand{\lig}{\texttt{row}\xspace}
\newcommand{\col}{\texttt{col}\xspace}
\newcommand{\start}{\texttt{start}\xspace}
\newcommand{\rownb}{\texttt{rownb}\xspace}

\newcommand{\spmv}{{\sc SpMV}\xspace}
\newcommand{\givaro}{{\sc Givaro}\xspace}
\newcommand{\linbox}{{\sc LinBox}\xspace}
\newcommand{\oski}{{\sc Oski}\xspace}

\newcommand{\F}{{\sf F}} 
\def\sigmabase{$\sigma$-basis~}
\def\so{{O {\;\!\tilde{}}\,}}

\renewcommand{\leq}{\leqslant}

\usepackage{paralist,array,enumerate,multirow,float}
\usepackage{color}

\urldef\bbjgdemail\url{\{Brice.Boyer,Jean-Guillaume.Dumas\}@imag.fr}
\urldef\pgemail\url{Pascal.Giorgi@lirmm.fr}

\begin{document}
\title{Exact Sparse Matrix-Vector Multiplication on GPU's and Multicore Architectures} 
\author{
Brice Boyer\footnote{\bbjgdemail, Universit\'e de Grenoble,
  Laboratoire Jean Kuntzmann, UMR CNRS 5224. 
  51, rue des Math\'ematiques, BP 53X, 38041
  Grenoble, France.}
\and Jean-Guillaume Dumas\footnotemark[1]~\thanks{Part of this work was
  done while the second author was visiting the Claude Shannon
  Institute and the University College Dublin, Ireland, under a CNRS
  grant.}
\and Pascal Giorgi\footnote{\pgemail, Universit\'e Montpellier 2, Laboratoire
  LIRMM, UMR CNRS 5506. F34095 Montpellier cedex 5,
  France.}
}
\maketitle
\begin{abstract}
  We propose different implementations of the sparse matrix--dense
  vector multiplication (\spmv{}) for finite fields and rings
  $\Zb/m\Zb$. We take advantage of graphic card processors (GPU) and
  multi-core architectures. 
  Our aim is to improve the speed of \spmv{} in the \linbox library,
  and henceforth the speed of its black box algorithms.
  Besides, we use this and a new parallelization of the
  sigma-basis algorithm in a parallel block Wiedemann rank
  implementation over finite fields.
\end{abstract}
%
%
\section{Introduction}
Nowadays, personal computers and laptops are often equip\-ped with
multicore architectures, 
as well as with more and more powerful graphic cards.
The latter ones can be easily programmable for a general purpose
computing usage (Nvidia Cuda, Ati Stream, OpenCL). 
Graphic processors offer nowadays quite frequently 
superior performance on a same budget as their CPU counterparts.
However, programmers can also efficiently use many-core CPUs for
parallelization e.g. with the OpenMP standard.

On the numerical side, several libraries automatically tune the sparse
matrix kernels \cite{Vuduc05oski,VuducM:05,Tvrdik:06} and recently
some kernels 
have been proposed e.g. for GPU's
\cite{ellr,buluc09,Bell:SpMV:SC:2009}.
In this paper we want to adapt those techniques for exact
computations and we first mostly focused on $\Zb/m\Zb$ rings, with $m$
smaller that a machine word.

The first idea is to use the numerical methods in an exact way as has
been done for dense matrix operations \cite{flas}. For sparse matrices,
however, the extraction of sparse matrices is slightly
different. Also, over small fields some more dedicated optimizations
(such as a separate format for ones and minus ones) can be useful.
Finally, we want to be able to use both multi-cores and GPU's at the
same time and the best format for a given matrix depends on the
underlying architecture.

Therefore, we propose an architecture with hybrid data formats,
user-specified or heuristically discovered dynamically.
The idea is that a given matrix will have different parts in different
formats adapted to its data or the resources. Also we present a
``just-in-time'' technique that allows to compile on the fly some
parts of the matrix vector product directly with the values of the
matrix. 

We have efficiently
implemented\footnote{\url{https://ljkforge.imag.fr/projects/ffspmvgpu/}}
``\texttt{Sp}arse \texttt{M}atrix-\texttt{V}ector multiplication''
(\spmv{}) on finite rings, together with the transpose product and
iterative process to compute the power of a matrix times a vector, or
a sequence of matrix products.

We also make use of this library to improve the efficiency of the
block Wiedemann algorithm's of the
\linbox\footnote{\url{http://linalg.org}} library.
Indeed, this kind of algorithm uses block ``black
box''~\cite{Kaltofen:1994:FHP} techniques:  
the core operation  is a matrix-vector multiplication and the matrix
is never modified. We use the new matrix-vector multiplication
library, together with a new parallel version of the sigma-basis
algorithm, used to compute minimal polynomials
\cite{Giorgi:2003:issac,jgd:2007:pasco}.

In section~\ref{sec:spmv} we present different approaches to the
parallelization of the \spmv operation, with the adaptation of
numerical libraries (section \ref{ssec:num}), new formats adapted to
small finite rings (section \ref{ssec:newf}) together with our new hybrid
strategy and their iterative versions (section \ref{ssec:block}).
Then in section~\ref{sec:rank} we propose a new parallelization of 
the block Wiedemann rank algorithm in \linbox, via the parallelization
of the matrix-sequence generation (section \ref{ssec:parseq}) and the
parallelization of the matrix minimal polynomial computation (section
\ref{ssec:parSB}).

\section{Sparse-Vector Matrix multiplication}\label{sec:spmv}
We begin with introducing some notations.
In this section, we will consider a matrix $A$; the element at row $i$, column $j$ is $A[i,j]$.
The number \nbnz is the number of non zeros elements in matrix $A$, it has \lig lines and  \col columns.
If $\xb$ and $\yb$ are vectors, then we perform here the operation $\yb \gets A\xb + \yb$.
The general operation $\yb \gets \alpha A\xb + \beta \yb$ can then be done by pre-multiplying $\xb$ and $\yb$ by $\alpha$ and $\beta$ respectively.
The \texttt{apply} operation in black box algorithms, or $\yb \gets A\xb$, is performed by first setting $\yb$ elements to zero.
For further use in block methods, we also provide the operation $\Yb \gets \alpha A \Xb + \beta \Yb$ where $\Xb$ and $\Yb$ are sets of vectors.
\subsection{Sparse Matrix Formats and Multiplication}\label{ssec:mul}
Sparse matrices arise form various domains and their shapes can
be very specific. 
Taking into consideration the structure of a sparse matrix can dramatically improve the performance of \spmv{}.
However, there is no general storage format that is efficient for all kind of sparse matrices.

Among the most important storage formats is the \coo (coordinate) format which stores triples.
It consists of three vectors of size \nbnz, named \data, \colid and \rowid, such that \texttt{data[k]=} $A$ \texttt{[rowid[k],colid[k]]}.

The \csr (compressed storage row) stores more efficiently the
previous representation: the \rowid field is replaced by a $(\lig+1)$ long \start vector such that if \texttt{start[$i$] $\leq k <$ start[$i+1$]}, then \texttt{data[$k$]$ = A $[$i$,colid[$k$]]}. 
In other words, \start indicates where a line starts and ends in the other two ordered fields. 

The \ell (ELLpack) format stores data in a denser way: it has \data and \colid fields such that  \texttt{data[$i,j_0$]$=A$[$i,$colid[$i,j_0$]]}, where $j_0$ varies between $0$ and the maximum number of non zero elements on a line of $A$.
One notices that these fields can be stored in row-major or column-major order.
A variant is the \ellr format that adds a $\lig$ long \rownb vector that indicates how many non zero entries there are per line.

The \dia (DIAgonal) is used to store matrices with non zero elements grouped along diagonals.
It stores these diagonals in an array along with offsets where they start.
We refer to~\cite{Bell:SpMV:SC:2009},\cite{ellr} for more details on these formats.

This very schematic description of a few well-known formats shows that each of them has pros and cons.
Our aim is to produce a more efficient implementation of the \spmv operation on finite fields than the one present in \linbox, taking first advantage of this variety of formats.
\subsection{Finite field representation}\label{ssec:ff}
We present now how the data is stored.
We use data types such as \texttt{float}, \texttt{int}.
Firstly, when doing modular linear algebra, we try to minimize the number of costly $fmod$ (reduction) operation calls.
For instance, we prefer if possible the left loop to the right one in the next figure:
\begin{multicols}{2}
\begin{verbatim}
for (i=0 ; i<n ; ++i){
    y += a[i] * b[i] ;
}
y = fmod(y,m);
\end{verbatim}
\begin{verbatim}
for (i=0 ; i<n ; ++i){
    y += a[i] * b[i] ;
    y = fmod(y,m);
}
\end{verbatim}
\end{multicols}
In this case, suppose $y=0$ and $a[i]$, $b[i]$ are reduced modulo $m$ at first, and $M$ is the largest representable integer.
Say that on $\Zb/m\Zb$, we represent the ring on $\oc0, m-1\fc$.
Then we can do at most $M/m^2$ such accumulations before reducing.
We can also represent the ring on $\left\llbracket-\left\lfloor\frac{m-1}{2}\right\rfloor,\left\lceil\frac{m-1}{2}\right\rceil\right\rrbracket$.
The latter representation enables us to perform twice more operations
before a reduction, but this reduction is slightly more expensive.
Another trade-off consists in choosing a \texttt{float} representation
instead of \texttt{double} (on the architectures that support
\texttt{double}).
Indeed, operations can be much faster on
\texttt{float} that on \texttt{double} but the \texttt{double}
representation lets us do more operations before reduction.
This is
particularly true on some GPU's.
\begin{figure}[H]
	\begin{center}
		\includegraphics[width=\textwidth]{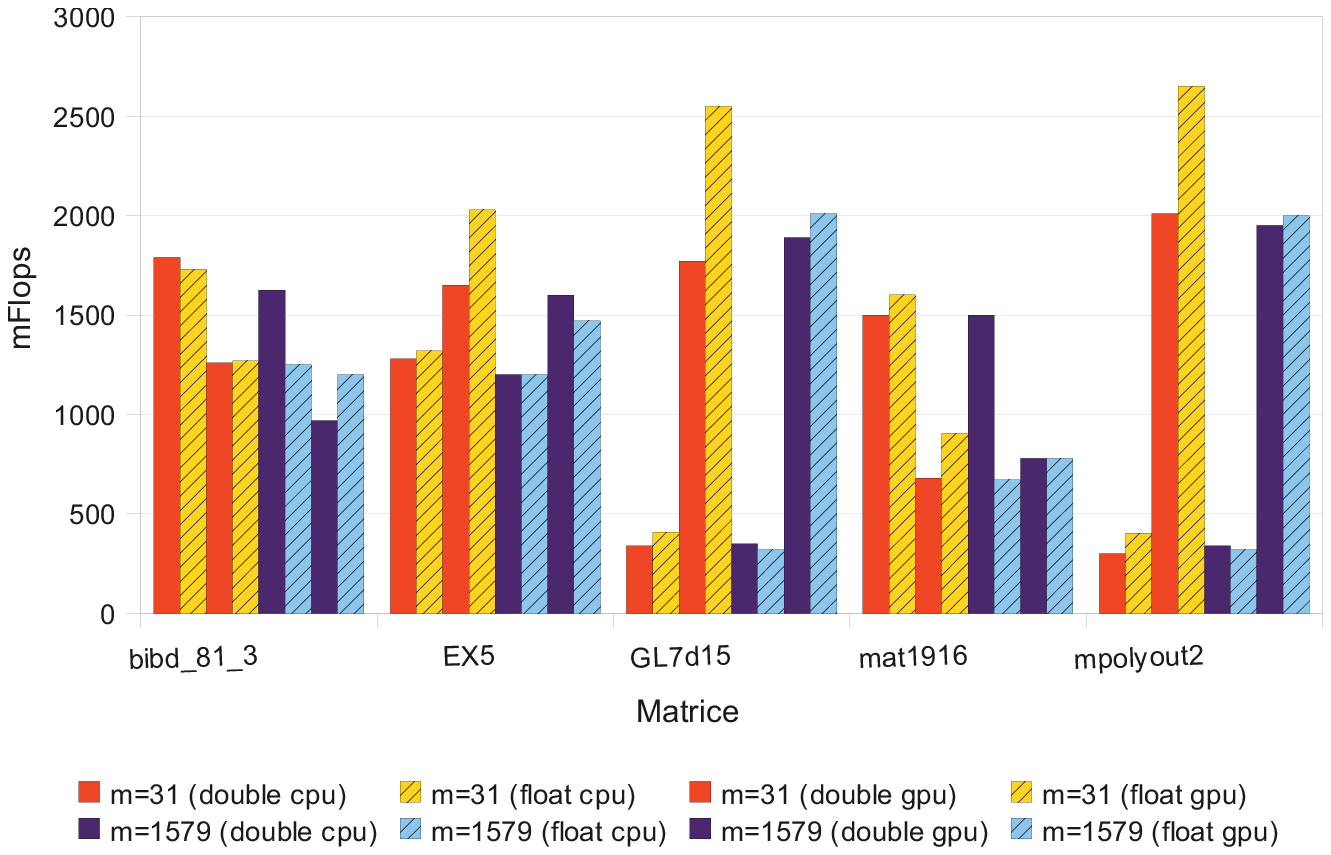}
	\end{center}
	\vspace{-1em}
	\caption{\texttt{float}--\texttt{double} trade-off for different sizes of $m$, on the CPU and GPU}
	\label{fig:dbl}
\end{figure}
In figure~\ref{fig:dbl}, we present variations in efficiency due to the storage data type and the size of $m$ on one core of a 3.2GHz Intel Xeon CPU and a Nvidia GTX280 GPU. 
The timings correspond to the average of 50 \spmv operations, where $\xb$ and $\yb$ are randomly generated on the CPU (which also takes into account every data transfer between the CPU and GPU).
The measures corresponds to the number of million floating point operations per seconds (flops); a \spmv operation requires $2*\nbnz$ such operations.
The performance correspond to the best ones achieved on the
matrices\footnote{matrices available
  at~\url{http://www-ljk.imag.fr/membres/Jean-Guillaume.Dumas/simc.html}} presented in table~\ref{tab:mats}.
\begin{table}[H]
	\centering
	{\small
	\newcolumntype{L}{>{\!\small}l<{\!}}
	\newcolumntype{C}{>{\!\small}c<{\!}}
	\begin{tabular}{C|LLLLL}
		name & mat1916   & bibd\_81\_3	& EX5   & GL7d15    & mpolyout2  \\
		\hline
		$\lig$& 1916		& 3240			& 6545	& 460261			& 2410560\\
		$\col$& 1916			& 85320			& 6545	& 171375		& 2086560 \\
		\nbnz	&195985	& 255960		& 295680&
                6080381			&15707520\\
                rank& 1916 & 3240 & 4740 & 132043 & 1352011\\
	\end{tabular}
	}
	\caption{Matrices overview}
	\label{tab:mats}
\end{table}
For further information about these techniques on these rings and fast arithmetic in
Galois extensions, see e.g.~\cite{flas}.
\subsection{Adapting numerical libraries}\label{ssec:num}
Another speed-up consists in using existing numerical libraries.
The ideas behind using them on the rings
$\Zb/m\Zb$, is twofold.
Firstly, we delay the modular reduction, secondly we can use highly optimized popular libraries and get instant speed-ups as compared to more naive self-written routines.

Just like BLAS libraries can be used to speed up modular linear algebra~\cite{Fflaflas}, we can use numerical libraries for our purposes, or get inspiration for our algorithms from their techniques.
For instance, there is the \oski library~\cite{Vuduc05oski} for sequential numerical \spmv{}, or the GPU implementation of \spmv{} by Nathan Bell \emph{et al.} in~\cite{Bell:SpMV:SC:2009}.
The BLAS specifications include sparse BLAS\footnote{\url{www.netlib.org/blas/blast-forum/chapter3.pdf}} but they are seldom fully implemented on free BLAS implementations. 

Unfortunately, they usually cannot be used as-is.
We need to extract submatrices from the sparse matrices, which
is more complicated than for its dense counterpart when the use of strides and dimensions suffices.
For instance, if one can do $b$ accumulations on $\yb[i]$ before reducing and the line $i$ of $A$ has $r_i$ non zero elements.
Then we want to split this line between $\left\lceil r_i/b\right\rceil$ matrices.
Finally, we can use the numerical libraries on these submatrices we have created.
The general algorithm reads like follows:
\begin{figure}[H]
\begin{verbatim}
spmv(y,A,x){
   foreach submatrix Ai in A do{
      spmv_num(y,Ai,x);
      reduce(y,m);
   }
}
\end{verbatim}
\vspace{-1em}
\caption{Using numerical routines}
\label{fig:num}
\end{figure}

\subsection{New formats}\label{ssec:newf}
Most of the formats implemented show a row-level parallelism, except \coo that has element-wise parallelism.
The \coo case is not obvious to implement and generally much slower.
The parallel efficiency of other formats will depend then on the length of the rows as well as the data regularity.
Unbalanced rows on a GPU architecture will produce many idle threads.
Two solutions exist: the vector approach of Bell (they split the rows into shorter chunks) or the rearranging of rows with permutations to sort the row according to their length.
The latter will not work in e.g. a power distribution of the row lengths.
The \ell format answers very well this problem because each row has the same length.

An other way to parallelize the \spmv operation is to split the matrix $A$ along rows to get smaller submatrices and treat them in parallel. 
We took this approach on the CPU \coo algorithm.

Also, we are dealing with large matrices, used many times as black-boxes.
Therefore there is a trade-off between the time spent on optimizing 
the matrix and how much faster these optimizations will make \spmv run.

Things to consider during preprocessing may include for instance:
reordering row-columns to create denser parts, choosing best-fitting
formats, cutting the matrix into efficient sub-matrices~(\cite{VuducM:05},\cite{Tvrdik:06})\dots
The preprocessing approach is taken by \oski: if the expected number of \spmv 
is very high, optimizing the matrix deeper will prove efficient.
\subsubsection{Base case: JIT}
One idea to improve \spmv on a given matrix is to hard code this
operation in a static library.
We read the matrix file and create a library that will apply this matrix to input vectors.
For instance the $\yb \gets \yb + A \xb$ operation on the matrix $
\begin{pmatrix}
	2 & 1 \\
	0 & 3
\end{pmatrix}$ would be translated to (if $m=27$)
\begin{verbatim}
void spmv(float * y, const float * x) {
   y[0] += 2*x[0] ;
   y[0] += x[1] ;
   y[0] = fmod(y[0],27);
   y[1] += 3*x[1] ;
   y[1] = fmod(y[1],27);
}
\end{verbatim}
Then we compile this generated file a as static library and use \texttt{dlopen} to access its functions.
As we can see in this example, one can implement various optimizations: rearranging the rows so that the work is more even (non implemented yet), replacing the occurrences of $\pm 1$ in the matrix by less costly additions or subtractions.
We have better control on what the compiler will produce.
However, large matrices take extremely long to compile.
\texttt{gcc} cannot compile the library if the source holds in one huge file, so we divide the matrix into parts of 1000 non zero elements and compile them.
Only then for instance, we could compile \texttt{bibd\_81\_3} but it takes 63s on the same Xeon machine.
Once it is compiled, the CPU version runs at 620 Mflops, which is
reasonably fast.

\subsubsection{Taking into account the $\pm1$}
The example of JIT and the observation that many matrices arising from different applications have a lot of $\pm 1_{F}$ tends to draw our attention to this special case. 
Moreover, many matrices on a small fields also share this property.
Thus we can extract two submatrices corresponding to the $1$ and $-1$ from the rest of the matrix and
replace multiplications by usually less expensive additions.
Besides, the \texttt{data} field in most formats (except \ell, \dia) can be forgotten as we
know they only consist of $1$ or $-1$: this reduces the memory usage.
Doing only additions as opposed to \texttt{axpy} also hugely delays reduction.
\begin{figure}[H]
	\begin{center}
		\includegraphics[width=0.7\textwidth]{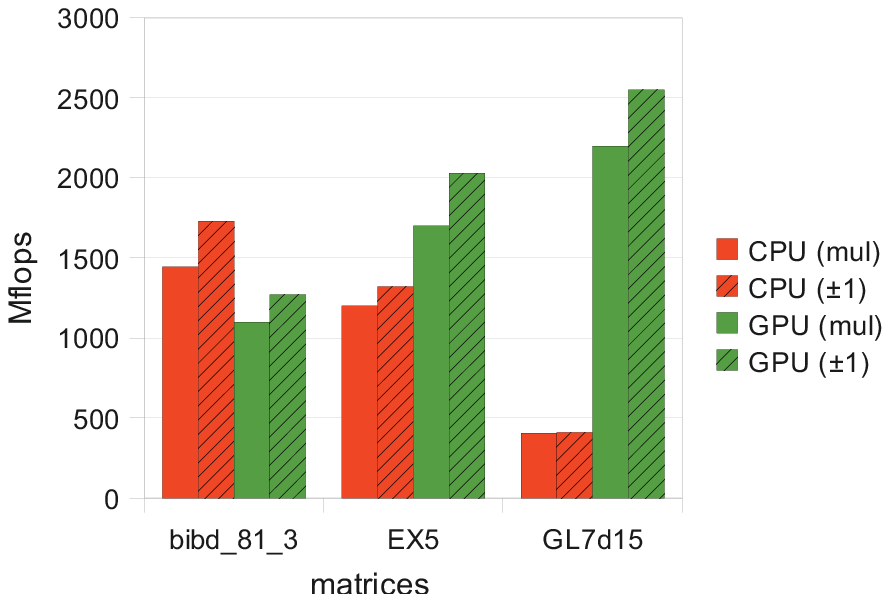}
	\end{center}
	\vspace{-1em}
	\caption{Speed improvement on one 3.2GHz Intel Xeon CPU and a Nvidia GTX280 GPU when segregating or not the $\pm1$}
	\label{fig:pm1}
\end{figure}
Figure~\ref{fig:pm1} shows a maximum 20\% improvement on a matrix with only 1s and 15\% on matrix with 50\% of $\pm1$.
\subsubsection{Basic Formats}
As evoked earlier, the matrix $A$ can be split into smaller submatrices.
These submatrices can have a format adapted to them and/or can be treated differently.
For instance, we can split row-wise and distribute these matrices for parallelism,
or split them column-wise as in the delaying case (figure~\ref{fig:num}.
This makes (possibly) many matrices that we each want to optimize individually
 so we get better overall performance.

We start with some observations.
The \coo format is slow due to the many
\texttt{fmod} calls, it is best used when the matrix is extremely
sparse. 
The \csr format is denser and can let delayed reduction occur,
but one has to ensure the row lengths are well balanced 
when parallelizing.
The \ell formats are very efficient on matrices that have roughly the same number of non zeros per line.
The \ellr format~(\cite{ellr}) is better for uneven rows lengths.
One difference in the CPU and GPU architecture makes the \ell row-major on the CPU (for better cache use) and column-major on the GPU (for better coalescing).
The following figure shows on one example (\texttt{bibd\_81\_3}) the variation of efficiency.
The data is normalized so that \csr is 1 on the CPU or GPU.
\begin{figure}[H]
\begin{center}
		\includegraphics[width=0.8\textwidth]{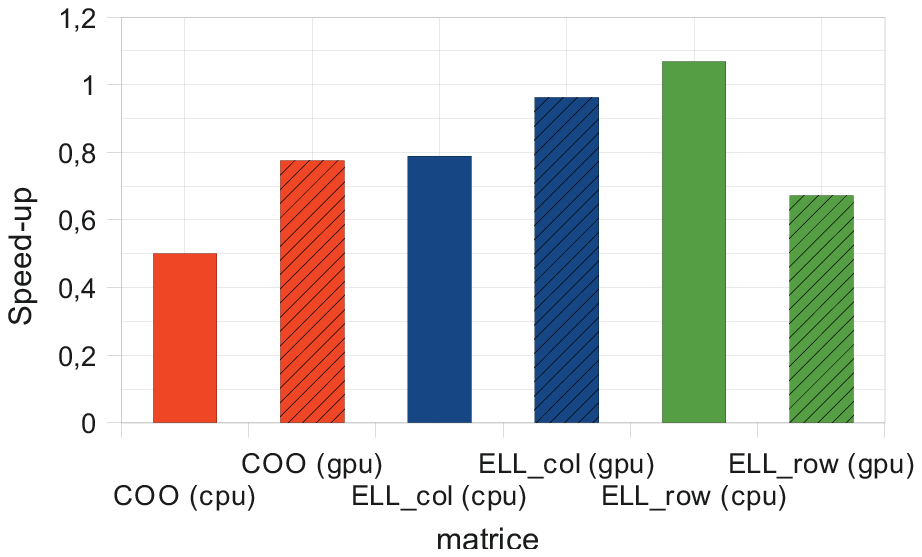}
	\end{center}
	\vspace{-1em}
	\caption{Speed-ups for various formats on matrix
          \texttt{bibd\_81\_3} both on one 3.2GHz Intel Xeon CPU and a Nvidia GTX280 GPU; reference
          is \csr on each architecture}
\end{figure}
\subsubsection{Hybridization}
The previous remarks lead us to combine these formats to take advantage of them.
A hybrid format such as \texttt{ELL(R)+COO} or \texttt{ELL(R)+CRS} leads to good performance on the GPU.
When the \ell part is taken out of a matrix, many rows can be
left empty.
Then, we use a format called \coos that is a
\csr format with pointers only to the non empty rows.
It has \data, \colid same as in \csr and \coo.
The number $\rowid[k]$ corresponds to the $k$\textsuperscript{th} non empty row that starts in \data and \colid at $\start[k]$.
This format could be avoided if we used row permutations and ordered the lines according to their weight.
\subsubsection{Heuristic format chooser}
The previous remarks show a great complexity in the formats and the cutting of the matrix.
We have implemented a user-helped heuristic format chooser.
For instance, the user can indicate if she wants to try and make use of $\pm1$.
If so, for each submatrix, the program tries to find an \emph{a priori} efficient format for them or if it fails, does not separate the $1$ or the $-1$ from the rest.
She can also indicate what is the format she wants to fill in priority.

The hybridization of the matrix is usually done as follows.
If the matrix is large enough and most of the lines are filled, it will try to fit a part of the matrix in an \ell or \ellr format.
This choice is supported by the observation that many matrices have a
$c+r$ row distribution where $c$ is some constant and $r \in \Zb$
varies and the fact that \ell is generally much faster that other formats
for matrices with even row weight.
The rest of the matrix will be put in a \csr, \coo or \coos format, according to the number of empty lines and the number of residual non zero elements.
Parameters that decide when segregating the $1$s, that choose the best
length for \ell matrix, etc., vary according to the architecture of
the computer and need some specific tuning. This tuning is not yet
provided at compile time but some of it could be automatically performed at
install time.

Experiments show that this heuristic often gives equal or better results that simple formats on the CPU and the GPU.

\subsection{Block and iterative versions}\label{ssec:block}
\subsubsection{Using multi-vectors}
We have described the \spmv operation $\yb\gets A\xb$ where $\xb$ and $\yb$ are vectors.
We also need $\xb$ and $\yb$ to be multi-vectors, for they may be used
for block algorithms.
There are at least two ways
of representing them : row or column-major order.
In the row-major order, we can use the standard \spmv many times (and align the vectors).
In the column-major order, we can write dedicated versions that try and make use of the cache.
Indeed, in this case, we traverse the matrix only once and $\xb$ and
$\yb$ are read/written contiguously.

\begin{figure}[H]
	\begin{center}
		\includegraphics[width=\textwidth]{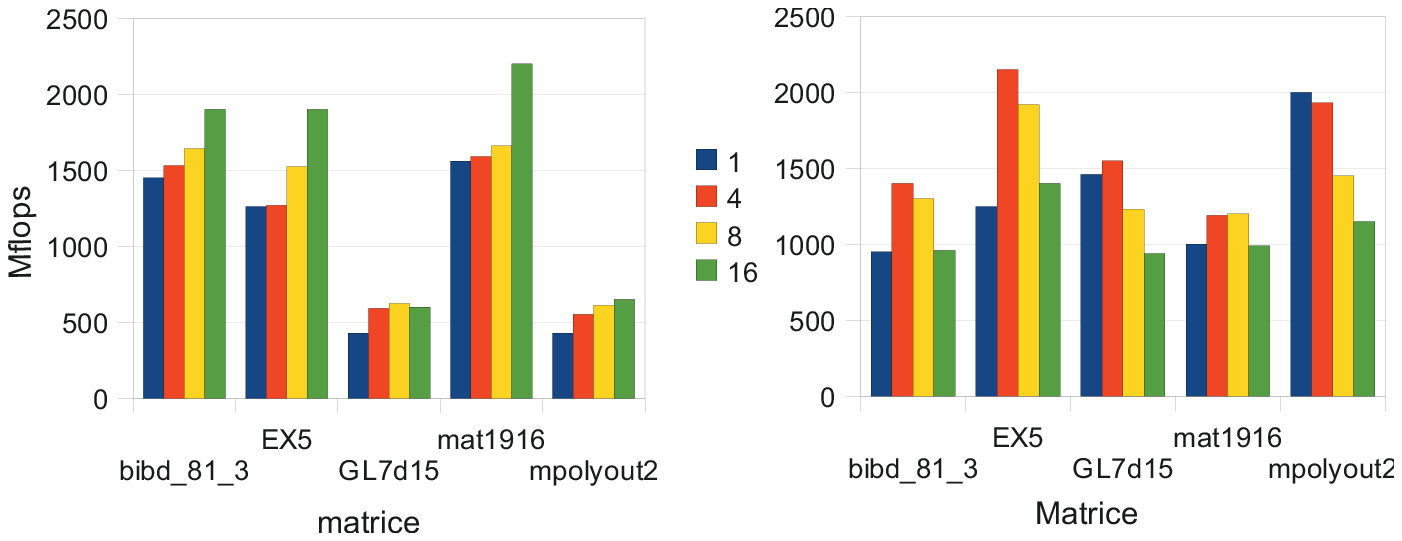}
	\end{center}
	\vspace{-1em}
	\caption{Matrix-multivector multiplication speed on one
          3.2GHz Intel Xeon CPU (left) and a Nvidia GTX280 GPU
          (right) for column-major multi-vectors, with $1,4,8$ and
          $16$ vectors.} 
	\label{fig:mv}
\end{figure}
On figure~\ref{fig:mv}, we note that on the CPU, using column-major
multivectors is a non negligible gain of speed.
On the contrary, the GPU implementation fails to sustain good
efficiency for blocks of more than 8 vectors and some large matrices start to
reach the memory limit.

\subsubsection{Performance issues}
The GPU operation on a single \spmv call from the host point of view is very slow because we need to move the vectors between the host and the device.
It is therefore only usable on operations that need no data moving between the host and the device. 
Examples include the computation $y\leftarrow~A^n~x$
or the computation of the sequence
$\left\{A^i x\right\}_{i\in \oc0,m\fc}$ that are used in many of the
black box methods.

\begin{figure}[H]
	\begin{center}
		\includegraphics[width=0.7\textwidth]{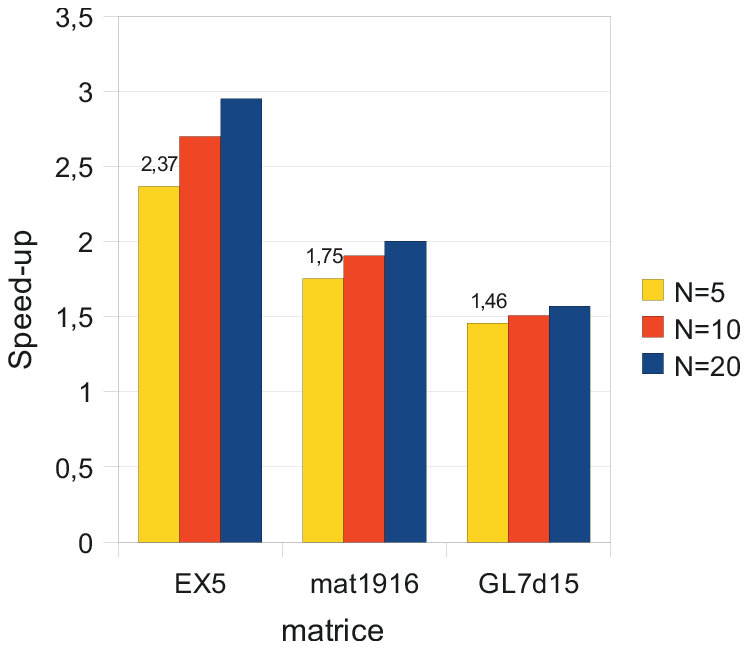}
	\end{center}
	\vspace{-1em}
	\caption{Nvidia GTX280 GPU speed up of $y \gets A^n x$ compared to $n$ times $y \gets A x$, with $n=5,10,20$}
	\label{fig:pow}
\end{figure}

On figure~\ref{fig:pow}, we illustrate this differences, mostly
reusing or not the data on the GPU, by comparing the performance of the following two pseudo-codes:
\begin{verbatim}
void smpv_n(y,A,x,n){
  y_d = copy_on_gpu(y);
  x_d = copy_on_gpu(x);
  A_d = copy_on_gpu(A);
  for (i=0 ; i<n ;++i) {
    y_d = A_d * x_d ; // spmv on GPU
    x_d = y_d;        // copy
  }
}
\end{verbatim}
\begin{verbatim}
void n_spmv(y,A,x,n){
  A_d = copy_on_gpu(A);
  for (i=0 ; i<n ;++i) {
    y_d = copy_on_gpu(y_i);
    x_d = copy_on_gpu(x_i);
    y_d = A_d * x_d ; // spmv on GPU
  }
}
\end{verbatim}
We confirm on figure~\ref{fig:pow} that it is highly desirable to
not move data on the GPU when avoidable.

\section{Parallel block Wiedemann algorithm}\label{sec:rank}

Some of the most representative applications requiring efficient sparse
matrix-vector product are 
blackbox methods based on the Lanczos/Krylov approach. In particular,
the method proposed by Wiedemann \cite{Wiedemann:1986:SSLE} and its
block version proposed by Coppersmith \cite{Coppersmith:1994:SHL} are
well suited to highlight efficiency of sparse matrix-vector product
since the latter is quite often their bottleneck.

As an application, we propose to improve the implementation of the Block Wiedemann rank algorithm presented in  \cite{jgd:2007:pasco}.
Let us first briefly recall the outline of this algorithm, we let the
reader refer to e.g. \cite{Turner:2006:BWR} for further details. \\

Let $A\in\F^{n\times n}$ be a matrix satisfying the preconditions of
\cite{Kaltofen:1991:SSLS}. Then the algorithm can be decomposed in three steps:
\begin{enumerate}
\item Compute the matrix sequence $S_i=  Y^T  A^i Y$ for $i=0..2n/s+O(1)$, with $Y\in\F^{n \times s}$ chosen at random
\item Compute the minimal matrix generator $F^A_Y\in\F^{s \times s}[x]$ of the matrix series $S(x)=\sum_i S_ix^i$
\item Return the rank $r=\deg(\det(F_{Y}^{A})) - \rm{codeg}   \det(F_{Y}^{A})$.
\end{enumerate}
Our approach is to separate the parallelization of each step.
The first step is clearly related to sparse matrix-vector product and we will re-use our tools presented in previous sections.
The second step needs the computation of a minimal matrix
generator. This can be achieved by a \sigmabase computation as
explained in \cite[section 2.2]{jgd:2007:pasco}. 
Finally, the last step reduces to computing the co-degree of the
determinant of the \sigmabase. 
The degree of the determinant being directly
computed as the sum of the row degrees of $F^A_Y$ since, due to the
\sigmabase properties, the matrix is
already in Popov form. 

\subsection{Parallelization of the matrix sequence
  generation}\label{ssec:parseq}

The parallelization proposed in \cite{jgd:2007:pasco} was to
ship independent set of vector blocks of $V$ to different cores and
apply them in parallel. Then gather the results to compute the dense
dot products by $U^T$.

An alternative is to use the \spmv{} library and let it take care of the
iteration with the algorithm of the preceding section.

In figure \ref{fig:seq} we compare both approaches:
\begin{figure}[H]
\begin{center}
		\includegraphics[width=0.7\textwidth]{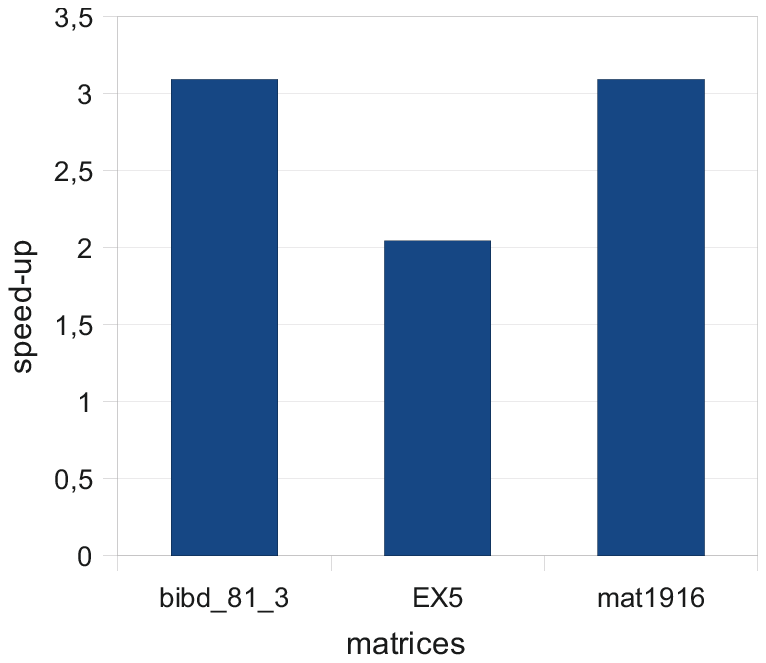}
	\end{center}
	\vspace{-1em}
\caption{Speed up from the new \spmv{} library compared to the native
  \linbox implementation in the generation of the matrix sequence
  ($2n$ iterations) on one core of a 2.33GHz Intel Xeon E5345 CPU}
\label{fig:seq}
\end{figure}
\subsection{Parallelization of the \sigmabase computation}\label{ssec:parSB}
One can efficiently compute \sigmabase using the algorithm {\sf PM-Basis} of \cite{Giorgi:2003:issac}. 
This algorithm mainly reduces to polynomial matrix
multiplication. Therefore a first parallelization approach 
  is to parallelize the polynomial multiplication.
 
\subsubsection{Parallel polynomial matrix multiplication}
Let $A,B \in \F^{n \times n}[x]$ be two polynomial matrices of degree
$d$. One can multiply $A$ and $B$ in $O(n^3d+n^2d\log d)$ operations in 
\F\, assuming \F\, has a $d$-th primitive root of unity
\cite{Cantor:1991:Kaltofen}. 
Assuming one has $k$ processors such that $k\leq n^2$, one can perform this multiplication with a parallel complexity of $O(\frac{n^3d}{k}+\frac{n^2d\log d}{k})$ operation in \F. Let us now see the sequential fast polynomial matrix multiplication algorithm and how it achieves such a parallel complexity:\\

\noindent {\it Fast Polynomial Matrix Multiplication:}\\
{\bf Inputs:}  $A,B \in \F^{n \times n}[x]$ of degree $d$, $\omega$ a $d$-th primitive root of unity inf \F.\\
{\bf Outputs:} $A\times B$\\
\hspace*{.4cm}1. $\bar{A} := DFT(A,[1,\omega,\omega^2,...,\omega^{2d}])$\\
\hspace*{.4cm}2. $\bar{B} := DFT(B,[1,\omega,\omega^2,...,\omega^{2d}])$\\
\hspace*{.4cm}3. $\bar{C}:= \bar{A}\otimes \bar{B}$\\
\hspace*{.4cm}4. $C:=\frac{1}{2d} DFT(\bar{C},[1,\omega^{-1},\omega^{-2},...,\omega^{-2d}])$\\
\hspace*{.4cm}{\bf return} C.\\

Here, $DFT(P,L)$ means the multi-points evaluation of the polynomial $P$ on each points of $L$, while $\otimes$ means the point-wise product.

\begin{itemize}
\item step 1,2 and 4 can be accomplished by using Fast Fourier Transform on each matrix entries which gives $n^2 \times O(d\log d)$ operations (see \cite[Theorem 8.15]{VonzurGathen:1999:MCA}). This clearly can be distributed on $k$ processors such that each processor achieves in  parallel the FFT on $\frac{n^2}{k}+O(1)$ matrix entries. This gives a parallel complexity of  $O(\frac{n^2d\log d}{k})$ operations in \F.
\item step 3 requires the computation of $2d$ independent matrix multiplications of dimension $n$, which gives $O(n^3d)$ operations in \F. One can easily see how to distribute this work on $k$ processors such that each processor has a workload  of $O(\frac{n^3d}{k})$ operations. \\
\end{itemize}

\begin{figure}[ht]
\includegraphics[width=0.5\textwidth]{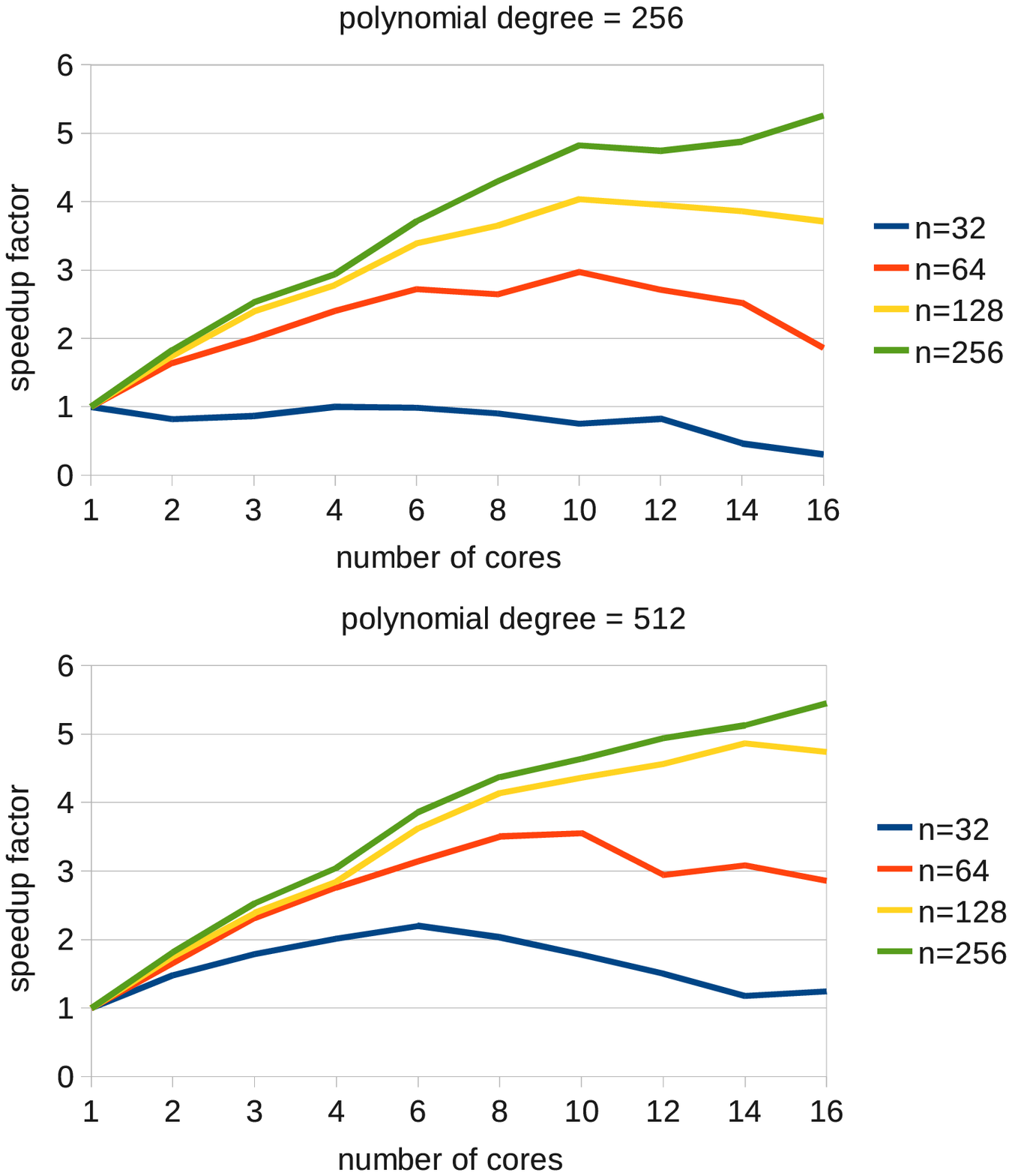}
\includegraphics[width=0.5\textwidth]{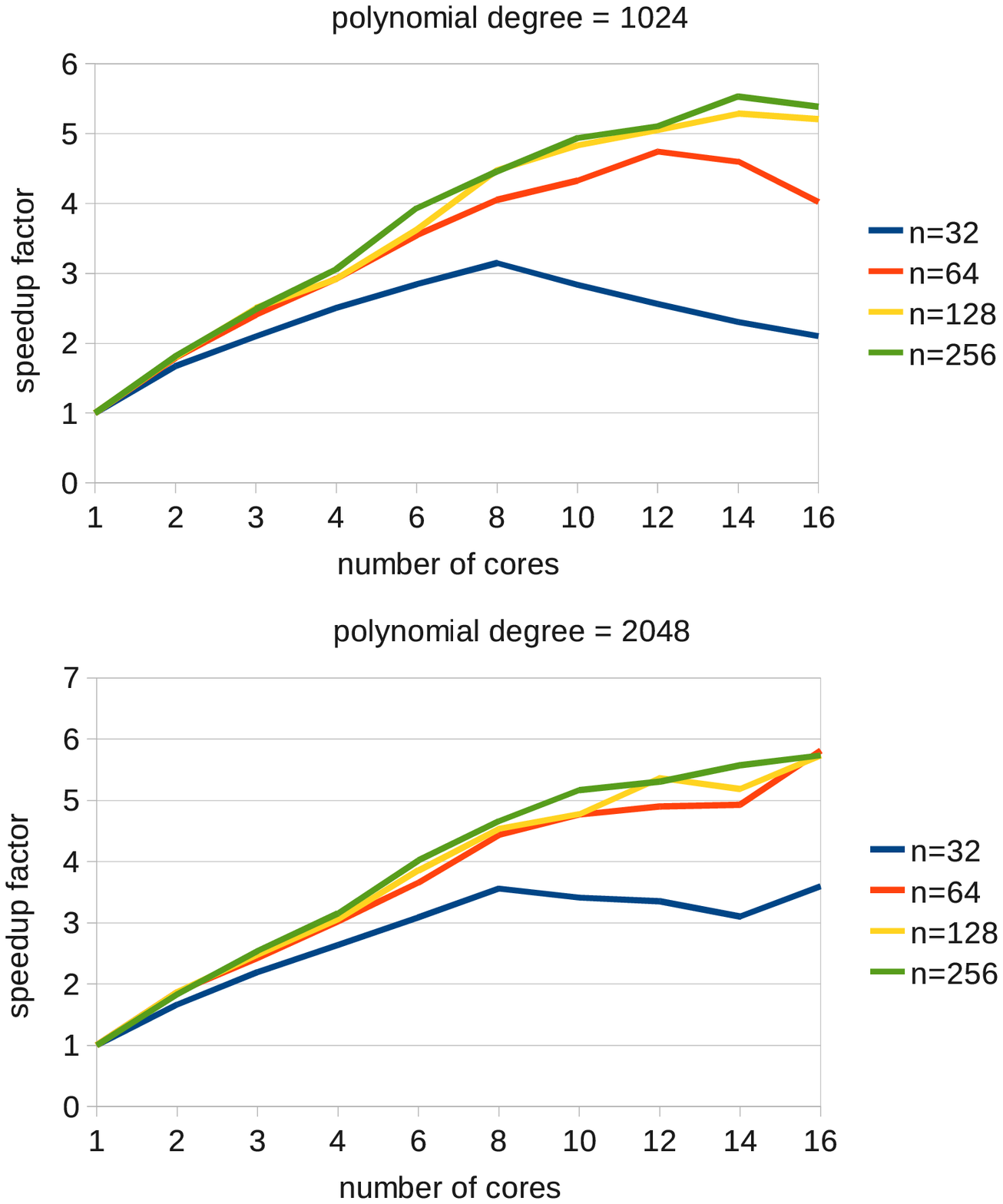}
\caption{Scalability of parallel polynomial matrix multiplication with LinBox and OpenMP on a 16 core machine (based on Quad-Core AMD Opteron). n is the matrix dimension.}\label{fig:multiplication}
\end{figure}

We report in figure \ref{fig:multiplication} the performance of the implementation of this parallel algorithm in the LinBox\footnote{www.linalg.org} library.
Our choice of using this parallel algorithm rather than another one, achieving a possible better parallel complexity, has been driven 
by the re-usability of efficient sequential components of the library (e.g. matrix multiplication) and the ease of use within the library itself 
(i.e. mostly the same code as sequential one, only some OpenMP pragmas
have been added). \\

One can see on figure \ref{fig:multiplication} that our coder does not completely match the theoretical parallel speedup.
The best we can achieved with 16 processors is a speedup of 5.5, which is only one third of the theoretical optimality. 
Nevertheless, one can see that with less processors (e.g. less than 4) the speedup factor is closer to 75\% of the optimality, which is quite fair.
We think this phenomenon can be explained by the underlying  many multi-core architecture (Quad-Core AMD Opteron), which may clearly suffers from cache
effect if computation are done on same chip or not.

As expected, we can also point out from figure \ref{fig:multiplication} that our implementation benefits at most from parallelism when matrices are larger.
Since workload on each core is more important, this allows to hide the penalty from memory operations and threads management of OpenMP. 
This remarks also applies on the degree but the impact is less important.
 
\subsubsection{Parallel \sigmabase implementation}\label{ssec:sigmabase}

According to the reduction of {\sf PM-Basis} to polynomial matrix multiplication, one can achieve a parallel 
complexity of $\so(\frac{n^3d}{k}+\frac{n^2d\log d}{k})$ operations in \F\, with $k$ processors for \sigmabase calculation, assuming $k\leq n^2$.
Therefore, it suffices to directly plug in our parallel polynomial matrix multiplication into the original code of the LinBox library to
get a parallel \sigmabase implementation.

\begin{figure}[ht]\center
\includegraphics[width=0.6\textwidth]{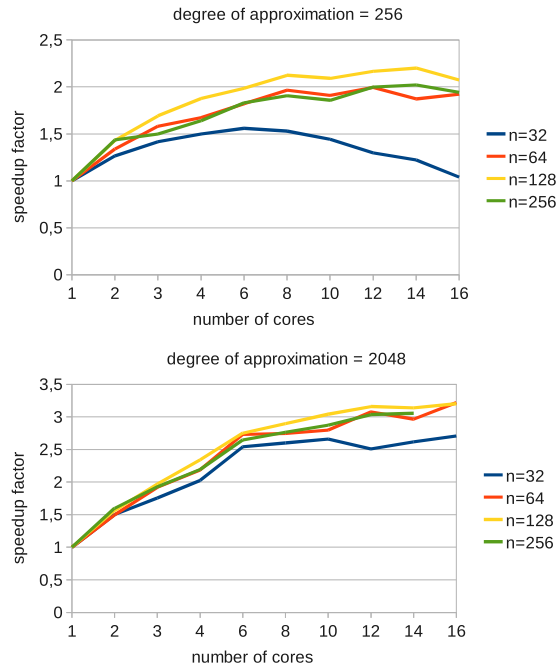}
\caption{Scalability of parallel \sigmabase computation with LinBox
  and OpenMP on a 16 core machine (based on a Quad-Core AMD Opteron). n is the matrix dimension of the series.}\label{fig:sigmabase}
\end{figure}

We report in figure \ref{fig:sigmabase} the performance of the parallel version of {\sf PM-Basis} algorithm within LinBox.
Here again, the speedup factor of parallelism is quite low when compared to the theoretical optimality.
At most we were able to obtain a speedup of 3 with 16 processors. 
However, this timings are consistent with the previous ones in figure \ref{fig:multiplication} where the best speedup was 5. 

One may notice that reduction to polynomial matrix multiplication of
the {\sf PM-Basis} algorithm relies on a divide a conquer approach on the
 degree of the approximation (see \cite[theorem
 2.4]{Giorgi:2003:issac}). Therefore, the recursion calls are made
 with smaller and smaller approximation's degrees, which leads to use
 less efficient parallel multiplications. Moreover, when the degree is
 too small the use of the {\sf M-Basis} algorithm of
 \cite{Giorgi:2003:issac} should be prefered since it becomes more
 efficient in practice. We have not yet implemented a parallel
 implementation of this algorithm in LinBox and this clearly
 affects the performance of our implementation.

\subsection{Parallel determinant co-degree}
Here we just launch in parallel the evaluations of the matrix
polynomial at different points, and the computation of the determinant of the
obtained matrix at the given point, and gather the results sequentially
with the \verb!Poly1CRT! class of \givaro.

\subsection{Parallel block Wiedemann performance}
In table \ref{tab:overall} we show the overall performance of our
algorithm on an
octo-processor Xeon E5345 CPU, 8 $\times$ 2.33GHz. 
\verb!*-LB! shows the timings of the current \linbox
implementation, where \verb!*-SpMV! presents our new improvement, both
in sequential and in parallel.
The speed-up for \spmv{} between 1 and 8 processors is slightly larger
than $5$ for all the matrices where the speed-up for \linbox ranges
from $4$ to $4.9$.
Furthermore, the speed-up obtained with \spmv{} versus \linbox on the
sequence generation seems scalable as it even
improves when used in a parallel setting.

\begin{table}[htb]\center
\begin{tabular}{|l||r|r|r|r|r|r|}
\hline
Matrix & \multicolumn{2}{c|}{mat1916} &   \multicolumn{2}{c|}{bibd\_81\_3}	& \multicolumn{2}{c|}{EX5}  \\
\hline
Cores & 1 & 8 & 1 & 8 & 1 & 8 \\
\hline
Seq-LB      & 15.09 & 3.08 & 47.73 & 12.41 & 84.21 & 20.22\\
Seq-\spmv{}     & 5.02 & 0.91 & 41.28 &  7.56 & 49.66 & 7.36\\
\hline
\sigmabase   & 9.02 & 1.64 & 18.45 &  3.63 & 37.45 & 8.39\\
Interpolation& 0.37 & 0.29 &  1.07 &  0.82 &  2.29 &  1.75\\
\hline
\hline
 Total-LB   & 24.48 & 5.01 & 67.25 & 16.86 &123.95 & 30.36\\
 Total-\spmv{} & 14.41 & \bf 2.84 & 60.80 & \bf 12.01 & 89.40 & \bf 17.50\\
\hline
\end{tabular}
\caption{Rank modulo 65521 with OpenMP Parallel block Wiedemann on a Xeon
  E5345, 8 $\times$ 2.33GHz}\label{tab:overall}
\end{table}

\section{Conclusion}
We have proposed a new \spmv library providing good results on
$\Zb/m\Zb$ rings.
To attain this efficiency it has been mandatory to augment the
complexity of the \spmv algorithms, since OpenMP, Cuda et al. all
manage differently the parallelization. Nonetheless, we provide new hybrid formats that improve the performance.
Moreover we have also specialized it to the computation of a sequence of
matrix-vector products together with a new parallelization of the
sigma-basis algorithm in order to enhance e.g. rank computations of
very large sparse matrices.

As seen in \ref{ssec:sigmabase}, a first parallelization of the
\sigmabase computation has been achieved. Its efficiency is not matching 
the expected scalability and lot of work needs to be done to circumvent this problem. First, a deeper study on the parallelization of 
\sigmabase computation has to be done.
Beside the parallelization of {\sf PM-Basis} and {\sf M-Basis}
algorithms themselves, we need to design new algorithms to avoid the
numerous task dependencies, 
inherent to the existing methods.
This will also enable an easier parallelization of early termination
strategies (requiring to interleave the generation sequence and the
\sigmabase computation).

Another important task is to extend the sigma-basis algorithm to work
on polynomial matrices over extension fields. 
Indeed the use of random projections $U$ and $V$ over extension fields might improve the probabilities to get the full
minimal polynomial of the matrix \cite{Kaltofen:1995:ACB,Gilles:study,Chen01efficientmatrix}. 
As shown in this paper and in \cite{jgd:2007:pasco}, \sigmabase needs only a polynomial matrix multiplication implementation to work.
In order to adapt current LinBox's implementation to extension field, we will use the same technique as \cite{flas}: first use Kronecker substitution to transform the extension field polynomial representation to an integer representation ; then use a Chinese remaindered version of the polynomial matrix multiplication to recover the resulting matrix polynomial over Z ; and finally convert back the integers using e.g. the REDQ inverse operation of \cite{jgd:2008:issac}.

The \spmv implementation also needs further work and other directions
to be explored.
For instance, we need to have dedicated implementations in $\Zb/2\Zb$ where $\xb$ and $\yb$ can be compressed.
More formats, including dense submatrices, have yet to be explored, which is linked to spending some more time on pre-processing the matrix: for instance the use of Metis\footnote{\url{http://glaros.dtc.umn.edu/gkhome/metis/metis/overview}} for partitioning and reordering $A$ would also improve the performance.
It will be interesting to deal with matrices such that $A$ and $\transpo{A}$ cannot be simultaneously stored~(\cite{buluc09}).
This problem indeed occurs on GPU's where on-chip memory is very limited.
Finally, we will also provide multi-GPU and hybrid GPU/CPU implementations.

\bibliographystyle{abbrv} 

\bibliography{spmv}
\end{document}